\documentclass{icrc}

\usepackage{times}
\usepackage{graphicx} 

\begin{document}

\title{Spectral cutoffs in EGRET gamma-ray sources}
\author[1,2]{O. Reimer}
\affil[1]{NRC/NASA/GSFC, Code 661, Greenbelt, MD 20771, USA}
\affil[2]{Now at: Institut f\"ur Theoretische Physik, Lehrstuhl
IV: Weltraum- \& Astrophysik, Ruhr-Universit\"at Bochum, D-44780 Bochum,
Germany}
\author[3]{D.L. Bertsch}
\affil[3]{NASA/GSFC, Code 661, Greenbelt, MD 20771, USA}

\correspondence{olr@egret.gsfc.nasa.gov}

\firstpage{1}
\pubyear{2001}


\maketitle

\begin{abstract}
The EGRET instrument has measured detailed photon spectra between 30 MeV and 
10 GeV, which are represented by means of single power-law fits for sources in
the 3EG catalog. However, various sources show indications of spectral cutoffs 
at GeV energies, which are poorly represented by such simple fits. In 
the case of well exposed or bright EGRET sources, a description of spectral 
cutoffs with more complicated functional forms appears to be applicable. An 
application for such multicomponent fits should be seen in extrapolations
beyond the energies accessible to EGRET, i.e. for detectability studies of 
low-threshold Imaging Atmospheric Cherenkov Telescopes (IACTs). In cases of 
unidentified gamma-ray sources positionally coincident with Supernova remnants, 
the spectral shape beyond power-law extrapolations might observationally explain 
why such prominent SNRs like $\gamma$ Cygni, IC 443 and CTA1 have not been detected 
during several observation campaigns performed by IACTs over the last years.
\end{abstract}

\section{Introduction}

Spectra of gamma-ray sources in the Third EGRET catalog (Hartman et al. 1999) 
are consistently represented by spectral indices from single power law fits. 
In cases of well measured source spectra (i.e. PSR J0633+1746 or PSR B0833-45), 
there are noticeable deviations from single power-law model fits reported 
\citep{fie97}. Here we apply higher-order spectral fits as introduced by 
\citet{ber00} to the dataset of "steady" unidentified gamma-ray sources used 
by \citet{geh00} for a gamma-ray source population study. This subset of the 3EG 
catalog sources has been used in order to discriminate different source classes 
among the population of unidentified gamma-ray sources, among other things by 
means of a comparison in the (simple power law) spectral index. It remains to be 
investigated, if differences in the average spectral index is a valid separation 
criteria in terms of the given uncertainties in the spectral fit as well as in
respect of more appropriate spectral description using higher order functional forms.
Additionally, satellite based gamma-ray measurements still do not connect to the higher
neighboring energies, currently accessible using ground-based Cherenkov telescopes. 
Therefore extrapolations need to be justified, either by appropriate source 
modeling and/or incorporating of additional observational information,
i.e. a quantitative indication of spectral breaks or cut-offs. Here we will
especially address cases of unidentified gamma-ray sources positionally coincident
with SNRs, as recently recompiled by \citet{rom99} and investigated at energies above 
300 GeV by \citet{buc98}.

\section{Analysis}

The "steady" unidentified sources from the 3EG have been fitted with three
functional forms, the standard single power law, two matching power laws, and
a power law with exponential cut-off:

\begin{equation}
\frac{\partial{J}}{\partial{E}}(E,K,E_{0},\lambda) = K \left(\frac{E}{E_{0}}\right)^{-\lambda}
\end{equation}

\begin{equation}
\frac{\partial{J}}{\partial{E}}(E,K,\lambda _{1}, \lambda _{2}) = \left\{ \begin{array}{ll} K \left(\frac{E}{1 \mathrm{GeV}}\right)^{-\lambda _{1}} (E \leq 1 \mathrm{GeV})  \nonumber\\
K \left(\frac{E}{1 \mathrm{GeV}}\right)^{-\lambda _{2}} (E \geq 1 \mathrm{GeV})
\end{array} \right.
\end{equation}

\begin{equation}
\frac{\partial{J}}{\partial{E}}(E,K,\lambda,E_{c}) = K \left(\frac{E}{300 \mathrm{MeV}}\right)^{-\lambda} exp\left(-\frac{E}{E_{c}}\right)
\end{equation}

$E_{0}$ has been set to the value determined by the EGRET spectral fitting routine
as used for the spectral fit given in the 3EG catalog. The location of the break 
energy has been set to 1000 MeV to keep the number of parameters at a minimum. 
With each fit, a reduced $\chi^{2}$ was obtained, followed by an F-Test to decide
if there is statistical justification for the transition from a single power law fit
to a higher order functional form. In the F-Test, a value of $p < 0.05$ is generally 
taken as a measure where the higher order fit is warranted. Summed data sets for
EGRET observations during CGRO observation cycles 1 to 4 were taken to maximize
the available statistics. Spectral studies of EGRET-detected blazars have been 
previously carried out by \citet{muk97}, \citet{poh97} and \citet{lin99}, but never 
beyond single power law representations. Because spectral hardening has been observed 
in gamma-ray blazars during outburst \citep{kum01}, an overall spectral comparison 
might not be similarly appropriate for AGN. They will be investigated later for 
different spectral representation than single power law, whenever possible on shorter 
timescales than the four years of CGRO observations used here. 

\section{Results} 

As already reported for the gamma-ray sources close to the Galactic plane in
\citet{ber00}, the majority of sources appears to be best represented by 
single power law fits. This trend has been confirmed for almost all unidentified 
sources at higher Galactic latitudes. However, only for significantly detected 
gamma-ray sources the question of departure from a single power law spectrum 
could be addressed with any chance of a statistically meaningful result.
Improvements from non-linear fitting are indicated by $\chi^{2}$ values lying 
right of the line in Fig.1 and Fig.2. In Fig.3 the choice of preference of either 
the two-power law fit or the power law with exponential cut-off is indicated by 
deviations from the line. Generally, a more complex spectra could only be established 
if sufficient counts have been observed, rather continuously distributed over 
the energy range accessible for EGRET, and obviously with bin entries above 1 GeV.\\
However, the F-Tests do not justify the statistical significance of each individual 
improvement in $\chi^{2}$. As indicated in Fig.4 (lines representing the F-Test 
probability $p = 0.05$), the majority of sources with better representation by 
non-linear fitting functions are insensitive to the form describing an spectral 
cuf-off within the currently 
available data (bottom-left quadrant in Fig.4). A clear indication of a statistically 
significant preference of either the two power law fit or the power law with 
exponential cutoff is only given for few individual gamma-ray sources: preference
of exponential cut-off are indicated in bottom-left quadrant, preference of 
two power law representation in top-left quadrant of Fig.4. 
Although many gamma-ray sources could not be compared at all to higher order
functional fits due to insufficient statistics above the chosen break or
cut-off energy, a tendency is indicated and will be probed by GLAST: The better
a gamma-ray source has been observed the more significant deviations from the
picture of single power fit representations are needed. EGRET seems to occupy 
a position where instrumental limitations generally do not satisfy more complex 
interpretations of measured source spectra, however well-measured individual 
gamma-ray sources already pointing towards structure in the gamma-ray source spectra 
unable to be adequately represented by a single power-law between 30 MeV and 
10 GeV.\\
Figure 5-8 give examples of individual unidentified sources with different functional
representations (Eq. 1-3). All these gamma-ray sources are positionally
coincident with SNRs, three of them suspected of resembling the properties
of neutron stars in gamma-rays (\citet{bra96}, \citet{bra98}, \citet{che98}).
These candidates are especially interesting because they were studied by ground-based 
Cherenkov telescopes, however yielding only upper limits so far \citep{buc98}.
For interpretation of these results the model of \citet{dru94} has been used
to extrapolate from the spectrum measured by EGRET. Here we suggest to extrapolate 
from EGRET observational results using better spectral representations \citep{pet00}. 
This might be an alternative or complementing way, especially under the premise that 
low-threshold Cherenkov telescopes will narrow the energetic gap to existing 
EGRET measurements.

\begin{figure}[t] 
\vspace*{2.0mm} 
\includegraphics[width=7cm]{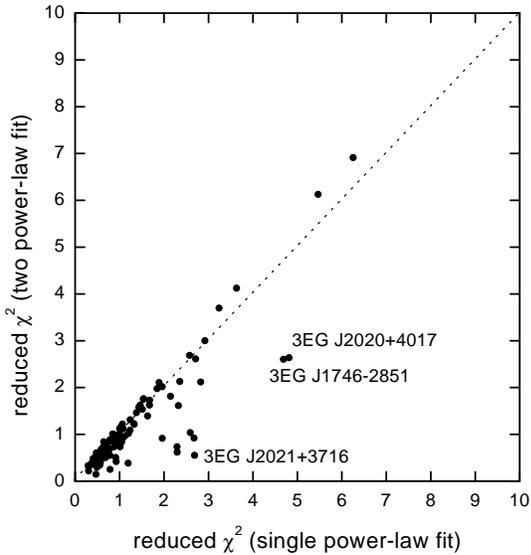}  
\caption{Comparison of the quality of a single power law fit vs. two power law
fit by means of reduced $\chi^{2}$}
\end{figure}
\begin{figure}[t] 
\vspace*{11.0mm} 
\includegraphics[width=7cm]{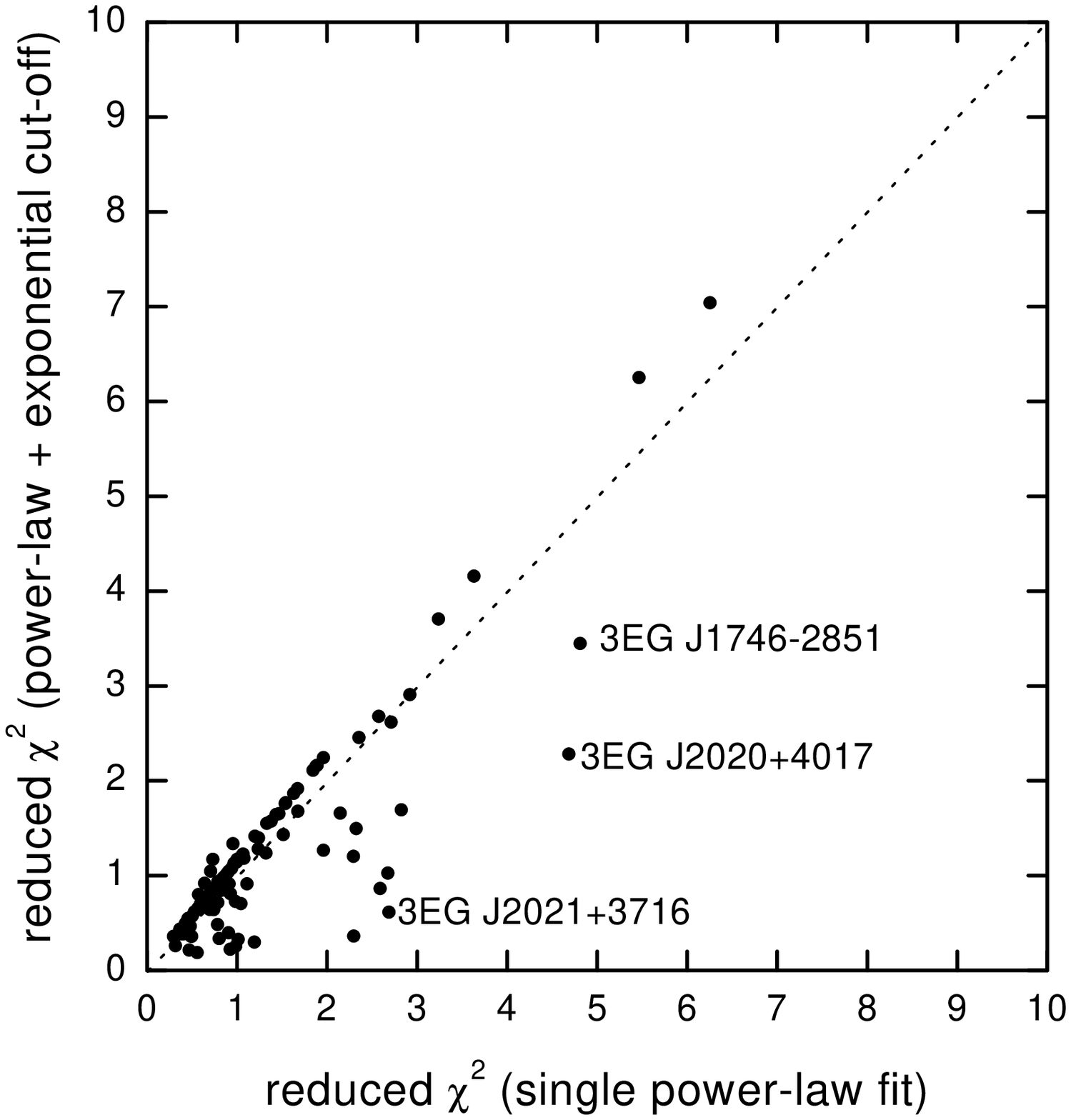}  
\caption{Comparison of the quality of a single power law fit vs. power law fit 
with exponential cut-off by means of reduced $\chi^{2}$}
\end{figure}
\begin{figure}[t] 
\vspace*{2.0mm} 
\includegraphics[width=7cm]{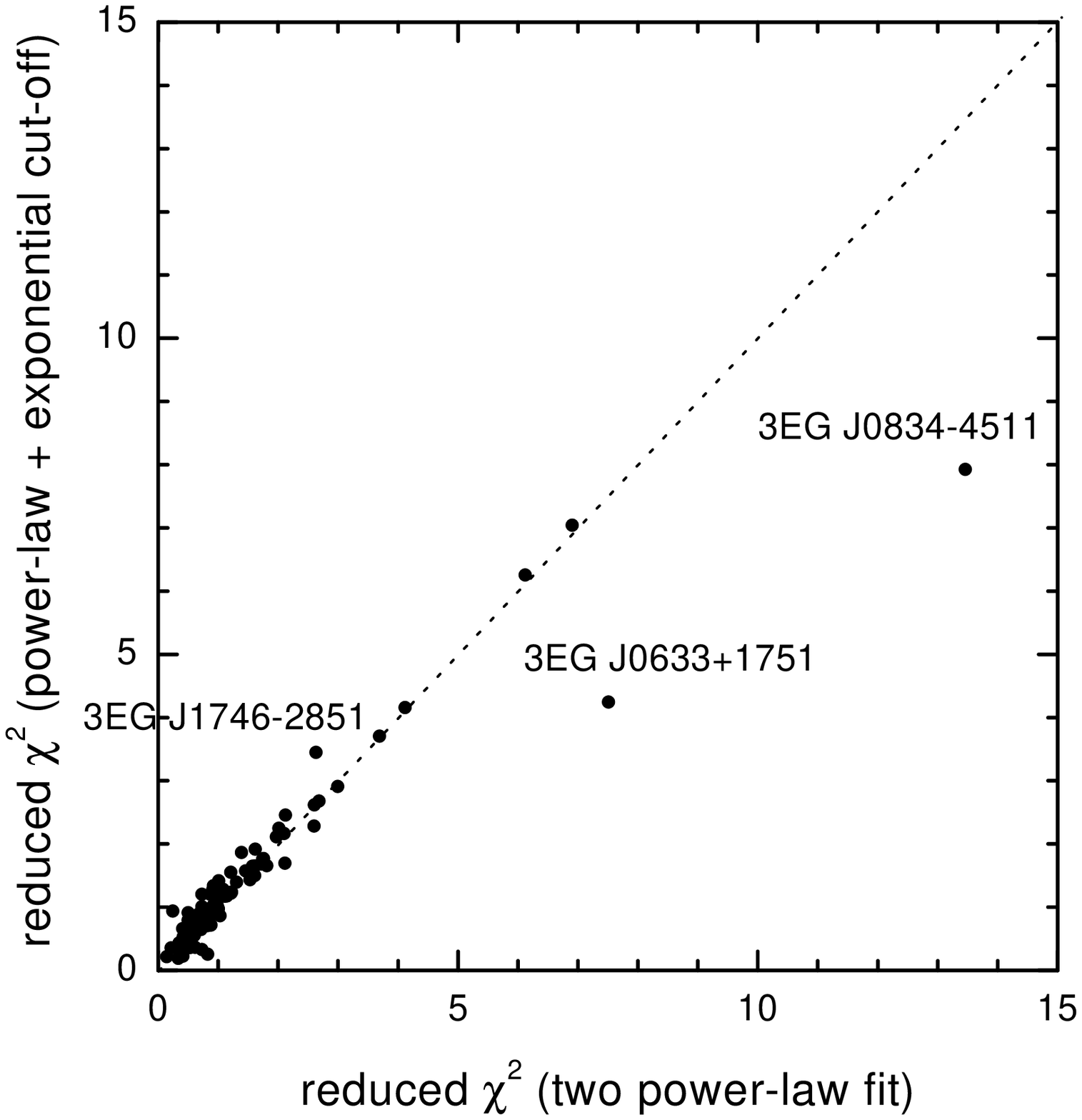}  
\caption{Comparison of the quality of a two power law fit vs. power law fit 
with exponential cut-off by means of reduced $\chi^{2}$}
\end{figure}
\begin{figure}[t] 
\vspace*{2.0mm} 
\includegraphics[width=8.3cm]{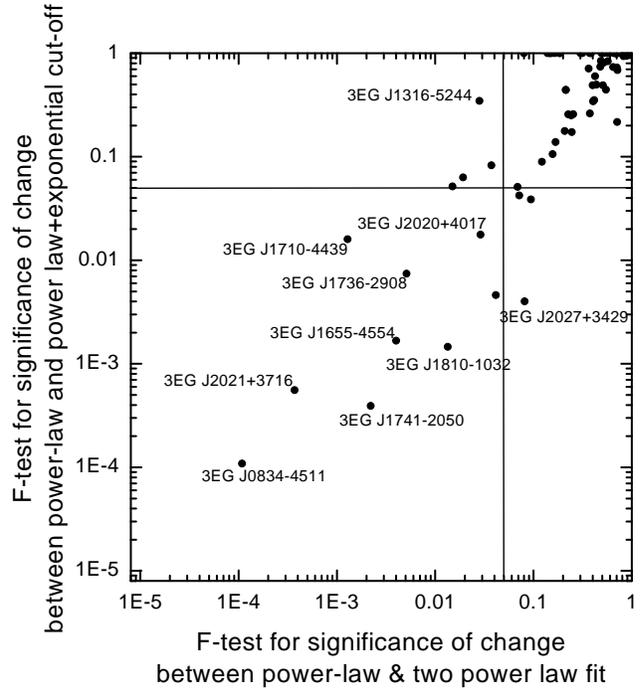}  
\caption{F-Test probabilities for the significance of the improvement obtained
by using higher order functional form for fitting the EGRET spectra. Probabilities
less than 0.05 indicate that the more complex fitting function is indeed statistically relevant.}
\end{figure}

%

%


\begin{acknowledgements}
O.R. acknowledges a NAS/NRC Associateship at NASA Goddard Space Flight Center.
\end{acknowledgements}

\begin{figure}[t] 
\vspace*{12.0mm} 
\includegraphics[width=7cm]{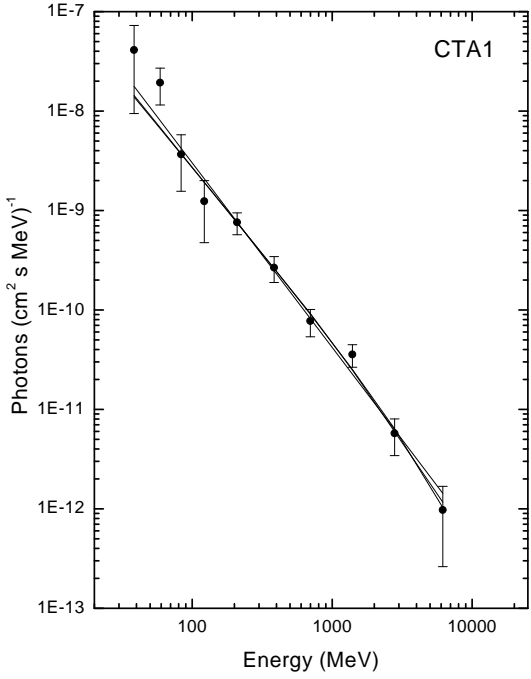}  
\caption{Photon energy spectrum of 3EG J0010+7309, positionally 
coincident with SNR G119.5+10.2 (CTA 1), shown with three different
representations of a spectral fit.}
\end{figure}
\begin{figure}[t] 
\vspace*{2.0mm} 
\includegraphics[width=7cm]{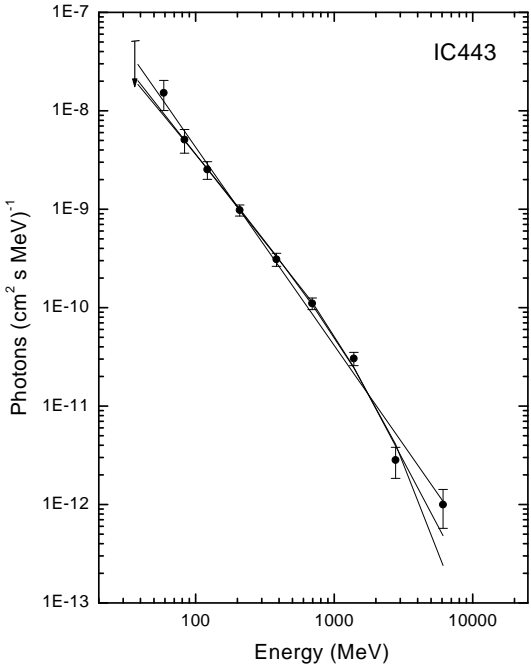}  
\caption{Photon energy spectrum of 3EG J0617+2238, positionally 
coincident with SNR G189.1+3.0 (IC443), shown with three different
representations of a spectral fit.}
\end{figure}
\begin{figure}[t] 
\vspace*{12.0mm} 
\includegraphics[width=7cm]{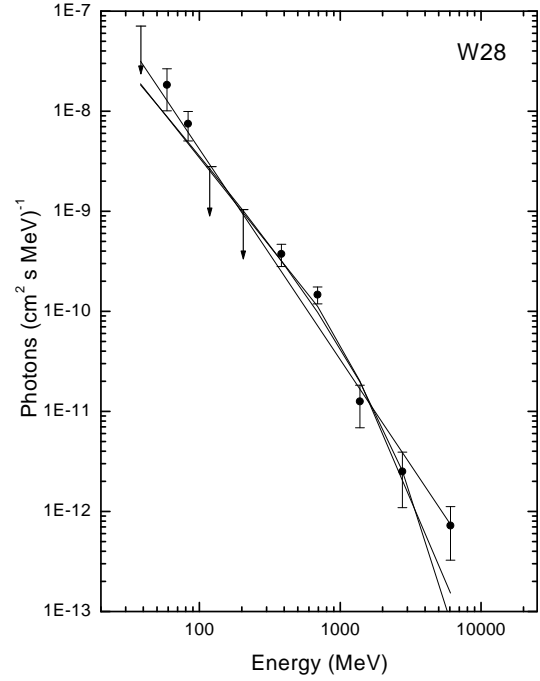}  
\caption{Photon energy spectrum of 3EG J1801-2312, positionally 
coincident with SNR G6.4-0.1 (W28), shown with three different
representations of a spectral fit.}
\end{figure}
\begin{figure}[t] 
\vspace*{2.0mm} 
\includegraphics[width=7cm]{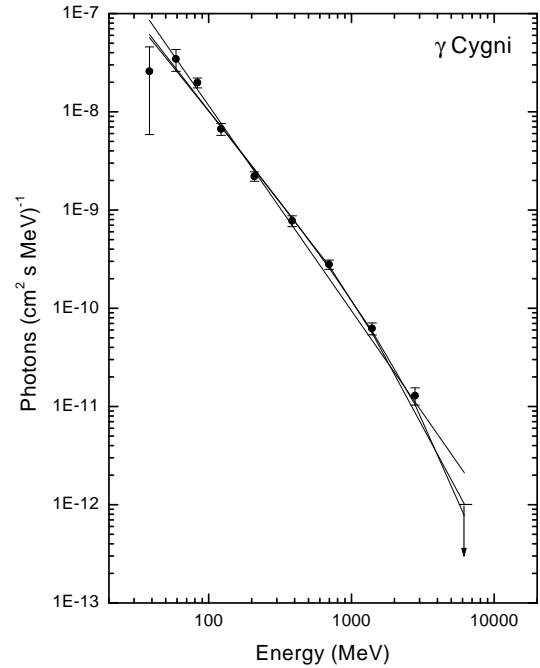}  
\caption{Photon energy spectrum of 3EG J2020+4017, positionally 
coincident with SNR G78.2+2.1 ($\gamma$ Cygni), shown with three different
representations of a spectral fit.}
\end{figure}

\end{document}